\documentclass[%
 showkeys,onecolumn,
 amsmath,amssymb,aps,prd
]{revtex4-2}
\usepackage{xcolor}
\usepackage[left, pagewise, edtable]{lineno}
\usepackage{graphicx}% Include figure files
\usepackage{dcolumn}% Align table columns on decimal point
\usepackage{bm}% bold math
\begin{document}

%\preprint{APS/123-QED}

\title{One-Loop Correction to the Casimir Energy in Lorentz-Violating $\phi^4$ Theory with Rough Membrane Boundaries}% Force line breaks with \\

\author{M. A. Valuyan}
 \altaffiliation[Also at ]{Energy and Sustainable Development Research Center, Semnan Branch, Islamic Azad University, Semnan, Iran}%Lines break automatically or can be forced with \\
\email{m-valuyan@sbu.ac.ir;ma.valouyan@iau.ac.ir}
\affiliation{%
 Department of Physics, Semnan Branch, Islamic Azad University, Semnan, Iran.%\\
}%

\date{\today}

\begin{abstract}
In this paper, we calculate the radiative correction to the Casimir energy for both massive and massless Lorentz-violating scalar fields confined between two membranes with rough surfaces in a 3+1 dimensional spacetime. The computations are performed for four types of boundary conditions: Dirichlet, Neumann, Periodic, and Mixed. A crucial element of our approach involves the use of position-dependent counterterms to incorporate the influence of boundaries within the renormalization program. To manage the divergences that emerge in the Casimir energy calculations, we apply the Box Subtraction Scheme (BSS) along with the cutoff regularization technique. We present and discuss results for various degrees of membrane roughness, emphasizing the consistency of our findings with theoretical expectations.
\end{abstract}

\keywords{Casimir Energy; Radiative Correction; Rough Membrane; Boundary Condition }
\maketitle
%\linenumbers
\section{\label{Intro.}Introduction}
\begin{figure}[th]\centering\includegraphics[width=6cm]{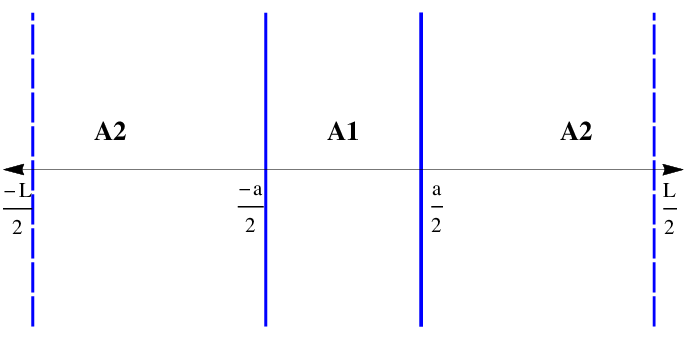}\hspace{2cm}\includegraphics[width=6cm]{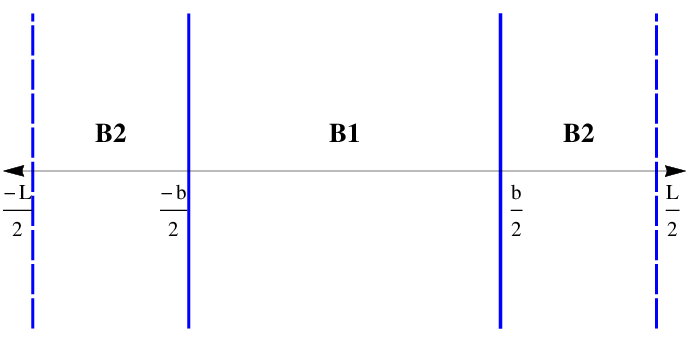}
\caption{\label{fig.BSS}
The left figure is ``A configuration'' and the right one is ``B configuration''.}
\end{figure}

The concept of Casimir energy, named after the Dutch physicist H. B. G. Casimir, originated in 1948 when he predicted the existence of an attractive force between two uncharged, perfectly conducting plates placed in a vacuum—now famously known as the Casimir effect~\cite{h.b.g.}. This force arises due to the quantized electromagnetic field between the plates, leading to a shift in the vacuum energy. Over the decades, Casimir's prediction has been experimentally validated~\cite{spaarnay,Lamoreaux1,Lamoreaux2,Mohideen,Bressi} and has since been extended to various fields, underscoring its fundamental significance in quantum field theory, condensed matter physics, astrophysics, and biophysics~\cite{other.works.1,other.works.2,bio.Casimir.1,bio.Casimir.2,bio.Casimir.3}. One contentious issue within the renormalization program, particularly concerning the radiative corrections of Casimir energy, is the choice of counterterms in the presence of boundary conditions or non-trivial backgrounds~\cite{EUR-Reza}. In most prior works, the renormalization program typically employs ``\emph{free counterterms},'' which are associated with a quantum field theory in Minkowski spacetime, unaffected by boundary conditions or external influences~\cite{free.counterterms.1,free.counterterms.2,free.counterterms.3,free.counterterms.4}. These counterterms are intrinsic to the underlying field theory, independent of any external factors such as boundaries or backgrounds. We argue, however, that when a quantum field is influenced by non-trivial boundary conditions or backgrounds, the renormalization program should account for these dominant conditions. Specifically, the breaking of translational symmetry should be reflected in the $n$-point function within renormalized perturbation theory\cite{mixed3D.MAN}. This perspective necessitates the use of position-dependent counterterms, as opposed to free counterterms, in the renormalization process. This approach has been thoroughly explained in previous studies \cite{JHEP.REZA}, where its physical implications and advantages have been extensively discussed. In this paper, we build upon this idea by employing position-dependent counterterms within the renormalization program to study the radiative correction to the Casimir energy for a Lorentz violating massive/massless scalar field confined by various boundary conditions—including Dirichlet, Neumann, Periodic, and Mixed Boundary Conditions (DBC/NBC/PBC/MBC)—between two parallel rough membranes. Incorporating Lorentz violation into the $\phi^4$ theory addresses key questions about potential deviations from standard physics frameworks, particularly in the realms of quantum gravity and high-energy physics~\cite{LV1.Series1,LV2.Series1}. Lorentz symmetry, a cornerstone of modern physics, is actively investigated for possible violations at high energies through both experimental and theoretical studies~\cite{LV1.Series2,LV2.Series2}. Studying the Casimir effect for scalar fields with Lorentz violation provides valuable insights into how such deviations influence boundary phenomena, including modifications to the energy spectrum and corrections to vacuum forces~\cite{LV1.Series3,Reza.LV}. Surface roughness is an unavoidable and critical factor in realistic scenarios, as it significantly affects the Casimir force. While idealized boundary conditions assume perfectly smooth surfaces, real surfaces inevitably deviate from this ideal, introducing fluctuations that alter the quantum vacuum energy. From a theoretical standpoint, understanding how roughness influences the Casimir effect is essential for refining models of quantum field interactions in realistic systems. Rough surfaces can notably impact the local density of quantum fluctuations, introducing corrections that are vital for achieving precise experimental predictions~\cite{Rough.refs1,Rough.refs2}. Practically, surface roughness is inevitable in real-world applications, especially in micro- and nano-scale systems where the Casimir effect is prominent. This sensitivity to roughness is particularly significant in the design and operation of microelectromechanical systems (MEMS) and nanotechnology devices, where unintended Casimir forces due to non-ideal surfaces can result in stiction, reduced functionality, or even device failure~\cite{Rough.refs3}. Incorporating roughness effects into theoretical models and experimental setups is therefore essential for optimizing the performance and reliability of these systems. It is important to note that both the zero- and first-order radiative corrections to the Casimir energy for multiple boundary conditions on two smooth parallel plates have been reported in previous works~\cite{Toms,MBCruz,Porfirio,Junior,Baron.1,Baron.2}. Additionally, the leading-order Casimir energy for a massless scalar field confined by DBC between two rough membranes in 2+1 dimensions has been documented in Refs. \cite{Rough.massless,Rough.2}. However, calculating the radiative correction to the Casimir energy for a self-interacting scalar field confined between two rough membranes under these boundary conditions is original and constitutes a novel contribution of this paper. A preliminary examination of the Casimir energy results obtained for first-order radiative corrections in 3+1 dimensions, as calculated between two perfectly smooth parallel planes in Refs.~\cite{Toms,MBCruz,Baron.1} and \cite{JHEP.REZA,EUR-Reza}, reveals discrepancies in both the magnitude and sometimes the sign of the energy. As noted in Ref.~\cite{EUR-Reza}, these differences arise from the type of renormalization program employed and the use of position-dependent counterterms. Specifically, the renormalization approach in~\cite{EUR-Reza}, which incorporates the effects of boundary conditions into the counterterms, establishes a self-consistent renormalization framework. This approach contrasts with other methods, such as those in~\cite{free.counterterm.5,free.counterterm.6}, which compute radiative corrections for systems with boundary conditions but retain free-space counterterms. The distinction in renormalization strategies leads to different results for the radiative corrections to the Casimir energy. Similar discrepancies are observed in works like~\cite{mixed3D.MAN,1D.Reza}, where position-dependent counterterms are also employed. In this paper, we adopt a position-dependent counterterm approach to eliminate the divergences associated with the bare parameters of the Lagrangian. As a result, our findings, even after eliminating membrane roughness, differ from those obtained through renormalization programs that employ free-space counterterms. However, when surface roughness effects are excluded, our results align with those reported in studies such as~\cite{JHEP.REZA,mixed3D.MAN,Ddim.MAN}, where the renormalization program incorporates position-dependent counterterms. In this study, we present a model to derive the Green's function in the presence of membrane roughness, enabling the computation of the Casimir energy. Since the Casimir energy is obtained by subtracting the vacuum energies of different configurations, sophisticated regularization methods are required to handle the infinities that arise in the calculations. Techniques such as zeta function regularization, dimensional regularization, the Box Subtraction Scheme (BSS), and the Green's function method have been developed and widely employed to extract finite, physically meaningful results from the formally infinite expressions encountered in these computations~\cite{methods.BSS,methods.ZetaF,Heat.kernel,other.regular.tech.1,other.regular.tech.3,other.regular.tech.4,other.regular.tech.5}. In this paper, we utilize the BSS along with a cutoff regularization technique to obtain the Casimir energy. In the well-known BSS approach, two comparable configurations are considered, and their vacuum energies are subtracted from each other under appropriate limiting conditions~\cite{BSS.2}. The BSS simplifies this subtraction process by introducing additional regulators, which help manage infinities more transparently and reduce the reliance on analytic continuation techniques. Specifically, a new definition for the Casimir energy is proposed, wherein the vacuum energies of two similar configurations are subtracted from each other. For instance, to obtain the Casimir energy for a quantum field confined between two parallel membranes separated by a distance $a$~(referred to as region $A1$ in Fig. (\ref{fig.BSS})), the following expression is used:
\begin{eqnarray}\label{BSS.DEF}
       E_{\mbox{\tiny Cas.}}&=&\lim_{L\to\infty}\lim_{b\to\infty}\Big[E^{(A)}_{\mbox{\tiny Vac.}}-E^{(B)}_{\mbox{\tiny Vac.}}\Big]
       \end{eqnarray}
where the expressions $E^{(A)}_{\mbox{\tiny Vac.}}=E^{(A1)}_{\mbox{\tiny Vac.}}+2E^{(A2)}_{\mbox{\tiny Vac.}}$ and  $E^{(B)}_{\mbox{\tiny Vac.}}=E^{(B1)}_{\mbox{\tiny Vac.}}+2E^{(B2)}_{\mbox{\tiny Vac.}}$ represent the total vacuum energies of the system $A$ and $B$ as depicted in Fig. (\ref{fig.BSS}). The limits $L,b\to\infty$ ensure that system $B$ asymptotically approaches the properties of Minkowski space. The parameters defining the system sizes in BSS, such as $a$, $b$, and $L$, act as regulators, providing a clear mathematical framework for the removal of infinities during the computation of the Casimir energy. One intriguing aspect of the Casimir effect is its sensitivity to the geometry and boundary conditions. While traditional studies often assume idealized, perfectly smooth surfaces, real-world boundaries invariably exhibit some degree of roughness. This roughness can significantly influence the Casimir force, introducing additional corrections that must be accounted for. As we have already mentioned, understanding these effects is crucial for the application of Casimir force measurements in practical scenarios, such as the design of micro- and nanoscale devices. In this paper, we explore the zero- and first-order radiative corrections to the Casimir energy within the framework of a Lorentz-violating $\phi^4$ theory, with a particular focus on the impact of rough membrane boundaries. Lorentz-violating theories, which deviate from the standard Lorentz invariance of spacetime, have garnered considerable interest due to their potential to explain various fundamental physical phenomena. The interplay between Lorentz violation and the Casimir effect offers a rich field for exploration, especially when considering realistic, rough boundary conditions. We present our calculations for four types of boundary conditions: Dirichlet, Neumann, Periodic, and Mixed Boundary Conditions. The structure of the paper is as follows: In Section \ref{Sec:model}, we outline the computational model and derive the vacuum energy at both zero- and first-order radiative corrections, highlighting the effects of Lorentz violation and boundary roughness. In the subsequent section, we present the zero-order Casimir energy and the first-order radiative correction for a massive/massless scalar field confined between two parallel smooth/rough membranes in 3+1 dimensions. Results for the four types of boundary conditions (DBC, NBC, PBC, and MBC) are detailed. Finally, in the concluding section, we summarize the findings and discuss the underlying physical principles driving the results.

\section{The model}\label{Sec:model}
The Lagrangian for a self-interacting scalar field, incorporating an {\ae}ther-like Lorentz-violating term, is expressed as~\cite{cruz.1,cruz.2,cruz.Fermion}:
\begin{eqnarray}\label{Lagrangian.}
  \mathcal{L}=\frac{1}{2}\Big[\partial_a\phi\partial^a\phi+\sigma_i (u_i\cdot\partial\phi)^2-m_0^2\phi^2\Big]-\frac{\lambda_0}{4!}\phi^4,
\end{eqnarray}
where the parameters $\lambda_0$ and $m_0$ represent the bare coupling constant and the bare mass of the real scalar field, respectively. Additionally, the coefficient $\sigma_i$ quantifies the scale of Lorentz symmetry breaking in the $i$-th direction of spacetime. This parameter is typically set to a value much smaller than one, and it encodes the Lorentz violation by coupling the derivative of the scalar field to a constant vector $u_i$. By varying the vector $u_i$, the direction of Lorentz violation can be oriented. The {\ae}ther vector $u_0 = (1,0,0,0)$ breaks Lorentz symmetry in the time direction. Similarly, the vectors $u_1 = (0,1,0,0)$, $u_2 = (0,0,1,0)$, and $u_3 = (0,0,0,1)$ correspond to Lorentz symmetry breaking in the spatial directions $x$, $y$, and $z$, respectively. We model the membrane as a thin layer in three spatial dimensions, defined by:
\begin{eqnarray}\label{Model.Membrane.}
       \frac{-L}{2}\leq x\leq\frac{L}{2},\hspace{2cm}\frac{-L}{2}\leq y\leq\frac{L}{2},\hspace{2cm}-\frac{a+h(x,y)}{2}\leq z\leq\frac{a+h(x,y)}{2},
\end{eqnarray}
where, $L^2$ represents the area of the membranes, and $a$ denotes the distance between them. The function $h(x,y)$ encodes the surface roughness of the membranes, with the assumption that $\text{Max}\{h(x,y)\} \ll a \ll L$. To simplify the expression of the membranes, we introduce the change of variables $x = v_1 L$, $y = v_2 L$, and $z = v_3 a \left[1 + \frac{h(x,y)}{a}\right]$. The valid domain for the new variables is $\frac{-1}{2} \leq v_1, v_2, v_3 \leq \frac{1}{2}$. Using the Lagrangian given in Eq.~(\ref{Lagrangian.}), the equation of motion for the free scalar field (with $\lambda_0 = 0$) is then obtained as:
\begin{eqnarray}\label{equation.O.motion}
     \Big[(1+\sigma_0)\partial_0^2-\mathbf{P}+m^2\Big]\phi=0,
\end{eqnarray}
where the operator $\mathbf{P}=\frac{1-\sigma_1}{L^2}\partial_1^2+\frac{1-\sigma_2}{L^2}\partial_2^2
     +\frac{1-\sigma_3}{a^2}[1+\mathcal{M}_a(v_1,v_2)]\partial_3^2$. Furthermore, the partial derivatives are expressed as $(\partial_0^2,\partial_1^2,\partial_2^2,\partial_3^2)=(\frac{\partial^2}{\partial t^2},\frac{\partial^2}{\partial v_1^2},\frac{\partial^2}{\partial v_2^2},\frac{\partial^2}{\partial v_3^2})$, and the function $\mathcal{M}_a(v_1,v_2)$ is given by $\mathcal{M}_a(v_1,v_2)=\frac{-2h}{a}+\frac{3h^2}{a^2}$. For membranes without any roughness (\emph{i.e.} $\mathcal{M}_a(v_1,v_2)=0$), the normalized wave function can be obtained by solving the following eigenvalue equation~\cite{Rough.2},
\begin{eqnarray}\label{Eigenvalue.0}
       \left[ \frac{-1+\sigma_1}{L^2}\partial_1^2+\frac{-1+\sigma_2}{L^2}\partial_2^2
     +\frac{-1+\sigma_3}{a^2}\partial_3^2\right]\phi^{(0)}=p^{(0)}\phi^{(0)}.
\end{eqnarray}
where $p^{(0)}$ denotes the eigenvalue corresponding to the eigenfunction $\phi^{(0)}$. In the following, we present the expressions for the wave function, associated eigenvalues, and Green's function for four types of boundary conditions (Dirichlet, Neumann, periodic, and mixed) simultaneously. For the case of DBC, we place two parallel membranes at $v_3=\pm1/2$, where the wave function must satisfy the condition $\phi^{(0)}(v_1,v_2,\pm1/2)=0$. In the case of NBC, the condition on the membranes located at \( v_3 = \pm 1/2 \) is given by \( \frac{\partial \phi^{(0)}}{\partial v_3} \big|_{v_3 = \pm 1/2} = 0 \). The MBC, on the other hand, simultaneously applies Dirichlet and Neumann boundary conditions. Specifically, for two parallel membranes, the DBC is applied to the left membrane (\emph{e.g.}, the membrane at \(v_3 = -1/2\)), while the NBC is applied to the right membrane (at \(v_3 = 1/2\)). It is important to note that reversing the order of the Dirichlet and Neumann boundary conditions on the membranes does not alter the vacuum energy expression. For the PBC, we use the condition $\phi^{(0)}(v_1,v_2,1/2)=\phi^{(0)}(v_1,v_2,-1/2)$. By solving the equation given in Eq. (\ref{Eigenvalue.0}) and applying the DBC/NBC/MBC/PBC to the quantum field at the membrane located at \( v_3 = \pm \frac{1}{2} \) (region $A1$ in Fig. (\ref{fig.BSS})), we obtain the normalized wave function as:
\begin{eqnarray}\label{phi0.solution}
         \phi^{(0)}_{n,\mbox{\tiny$\mathcal{B}$}}(v_1,v_2,v_3)=\sqrt{2}e^{i k_1.v_1}e^{i k_2.v_2}\mathcal{H}_{n,\mbox{\tiny$\mathcal{B}$}}(v_3)=\sqrt{2}e^{i k_1.v_1}e^{i k_2.v_2}\times\left\{
         \begin{array}{c}
           \sin\Big(k_{n,\mbox{\tiny$\mathcal{B}$}}(v_3+\mbox{\small$\frac{1}{2}$})\Big)\hspace{1cm}\mbox{\small$\mathcal{B}=\{\mathcal{D},\mathcal{M}\}$}, \\
           \cos\Big(k_{n,\mbox{\tiny$\mathcal{B}$}}(v_3+\mbox{\small$\frac{1}{2}$})\Big)\hspace{1cm}\mbox{\small$\mathcal{B}=\{\mathcal{N},\mathcal{P}\}$}.
         \end{array}\right.
\end{eqnarray}
The subscript \mbox{\small$\mathcal{B} = \{\mathcal{D}, \mathcal{N}, \mathcal{M}, \mathcal{P}\}$} denotes the type of boundary condition. Specifically, \mbox{\small$\mathcal{D}$}, \mbox{\small$\mathcal{N}$}, \mbox{\small$\mathcal{M}$}, and \mbox{\small$\mathcal{P}$} represent DBC, NBC, MBC, and PBC, respectively. For Dirichlet and NBC, the wavenumber \(k_{n, \text{\tiny$\mathcal{B}$}}\) is given by \(n \pi\), while for MBC, it is \(k_{n, \text{\tiny$\mathcal{M}$}} = \left(n + \frac{1}{2}\right) \pi\). In the case of Dirichlet and NBC, the allowed values for \(n\) are \(n \in \mathbb{N}\), whereas for MBC, \(n \in \mathbb{N} \cup \{0\}\). In the case of PBC, the allowed wavenumber is $k_{n,\mbox{\tiny\(\mathcal{P}\)}}=2n\pi$, where $n=\pm1,\pm2,\pm3,...$. The corresponding eigenvalue, obtained from Eq.~(\ref{Eigenvalue.0}), is:
\begin{eqnarray}\label{eigenvalue.0.D&M}
          p^{(0)}_{n,\mbox{\tiny$\mathcal{B}$}} = \frac{1}{L^2} \left[\left.(1-\sigma_1)k_1^2 + (1-\sigma_2)k_2^2\right.\right] +\frac{1-\sigma_3}{a^2} k_{n,\mathcal{B}}^2.
\end{eqnarray}
To account for the roughness properties of the membranes in the eigenvalue, we use a perturbative approach \cite{Rough.massless,Rough.2}. This results in the eigenvalue being expressed as follows:
\begin{eqnarray}\label{First.Order.Eigenvalue}
        p^{(1)}_{n,\mbox{\tiny$\mathcal{B}$}}=\frac{1-\sigma_3}{a^2}k_{n,\mathcal{B}}^2\int_{\frac{-1}{2}}^{\frac{1}{2}}\int_{\frac{-1}{2}}^{\frac{1}{2}}dv_1dv_2\mathcal{M}_a(v_1,v_2)
        =\frac{1-\sigma_3}{a^2}k_{n,\mathcal{B}}^2 M.
\end{eqnarray}
Using Eq.~(\ref{equation.O.motion}), the wavenumber equation can now be written as:
\begin{eqnarray}\label{wavenumber}
        \omega^2_{n,\mathcal{B}}=\frac{1}{1+\sigma_0}\left[p^{(0)}_{n,\mbox{\tiny$\mathcal{B}$}}+p^{(1)}_{n,\mbox{\tiny$\mathcal{B}$}}+m^2\right].
\end{eqnarray}
In the literature, the Green's function expression for the massive scalar field confined between two smooth parallel plates by Dirichlet, Neumann, Periodic and MBC with area $L^2$ and distance $a$ after the Wick rotation have been known~\cite{LP.MAN}. This standard form for the case of DBC/NBC/PBC is commonly written as:
\begin{eqnarray}\label{Greens.function.withoutLV&Roughness}
      G_{\mbox{\tiny$\mathcal{B}$}}(a;v,v')=\frac{2}{a}\int\frac{d\omega}{2\pi}\int\frac{d^2\mathbf{k}}{(2\pi L)^2}\sum_{n_{\mathcal{B}}}\frac{e^{-\omega(t-t')}
      e^{i\mathbf{k}.(\mathbf{v}-\mathbf{v}')}\mathcal{H}_{n,\mbox{\tiny$\mathcal{B}$}}(v_3)\mathcal{H}_{n,\mbox{\tiny$\mathcal{B}$}}(v'_3)}
      {\omega^2+\frac{k_1^2+k_2^2}{L^2}+\frac{k_{n,\mathcal{B}}^2}{a^2}+m^2},
\end{eqnarray}
here \mbox{\small$\mathcal{B}=\{\mathcal{D},\mathcal{N},\mathcal{P}\}$}, $\mathbf{v}=(v_1,v_2)$, and $\mathbf{k}=(k_1,k_2)$. Furthermore, the expression for the function \(\mathcal{H}_{n,\mbox{\tiny$\mathcal{B}$}}(v_3)\) is introduced in Eq.~(\ref{phi0.solution}). This form for MBC has been reported as \cite{mixed3D.MAN}:
\begin{eqnarray}\label{Greens.function.MBC.without.LV&Roughness}
          G_{\mbox{\tiny $\mathcal{M}$}}(a;v,v')=\frac{1}{a}\int\frac{d\omega}{2\pi}\int\frac{d^2\mathbf{k}}{(2\pi L)^2}\sum_{n=0}^{\infty}\frac{
          \begin{array}{c}
            e^{-\omega(t-t')}e^{i\mathbf{k}\cdot (\mathbf{v}-\mathbf{v}')}[\sin(k_{n,\mathcal{M}}v_3)+(-1)^n\cos(k_{n,\mathcal{M}}v_3)] \\
            \times[\sin(k_{n,\mathcal{M}}v'_3)+(-1)^n\cos(k_{n,\mathcal{M}}v'_3)]
          \end{array}}{\omega^2+\frac{k_1^2+k_2^2}{L^2}+\frac{k_{n,\mbox{\tiny$\mathcal{M}$}}^2}{a^2}+m^2}.
\end{eqnarray}
The roughness of the membranes, combined with Lorentz violation in the scalar field, modifies the standard forms of the Green's function expressions. To address these modifications, we provide a detailed derivation of the new Green's function expressions for all types of boundary conditions in Appendix \ref{Appendix.GreensFunc.}. The resulting Green's function, accounting for membrane roughness and Lorentz violation under both boundary conditions, is given by:
\begin{eqnarray}\label{Greens.function.DBC.LV&Roughness}
         \tilde{G}_{\mbox{\tiny$\mathcal{B}$}}(a;v,v')=\frac{G_{\mbox{\tiny$\mathcal{B}$}}(\tilde{a}_1;v,v')}{\sqrt{(1+\sigma_0)(1-\sigma_1)(1-\sigma_2)}}
\end{eqnarray}
where $\tilde{G}_{\mbox{\tiny$\mathcal{B}$}}(a;v,v')$ denotes the modified Green's function expression, which incorporates both the roughness properties of the membranes and the Lorentz violation of the quantum field. Furthermore, the parameter $\tilde{a}_1=\frac{a}{\sqrt{1-\sigma_3}\sqrt{1+\mathcal{M}_a(0,0)}}$.
\par To calculate the vacuum energy up to the first-order in the coupling constant $\lambda$ based on the $\phi^4$ theory described by the Lagrangian in Eq. (\ref{Lagrangian.}), the bare parameters $m_0$ and $\lambda_0$ must be renormalized. This is achieved by introducing a rescaling of the scalar field with the parameter $Z=\delta_z+1$, commonly known as the field strength parameter. This rescaling modifies the Lagrangian given in Eq. (\ref{Lagrangian.}) to:
      \begin{eqnarray}\label{lagrangian.renormalized}
     \mathcal{L}&=&\frac{1}{2}(\partial_\mu\phi_r)^2+\frac{1}{2}\sigma_i(u_i\cdot\partial\phi_r)^2-\frac{1}{2}m^2\phi_r^2-\frac{\lambda}{4!}\phi_r^4\nonumber\\
     &&+\frac{1}{2}\delta_Z(\partial_\mu\phi_r)^2+\frac{1}{2}\delta_Z\sigma_i(u_i\cdot\partial\phi_r)^2-\frac{1}{2}\delta_m\phi^2-\frac{\delta_{\lambda}}{4!}\phi_r^4,
\end{eqnarray}
where $\delta_m=Zm_0^2-m^2$ and $\delta_\lambda=Z^2\lambda_0-\lambda$ denote the mass and coupling constant counterterms, respectively. The Feynman rules corresponding to these counterterms are defined as follows:
\begin{eqnarray}\label{Fynmann.Rule.Counterterms}
   \raisebox{0mm}{\includegraphics[width=1cm]{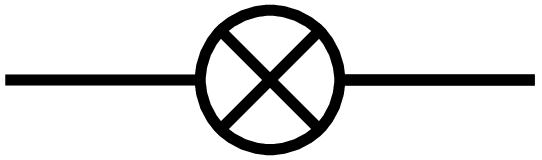}}&=&i\big[(p^2+\sigma_i(u_i\cdot p)^2)\delta_Z-\delta_m\big],\nonumber\\
   \raisebox{-2mm}{\includegraphics[width=0.7cm]{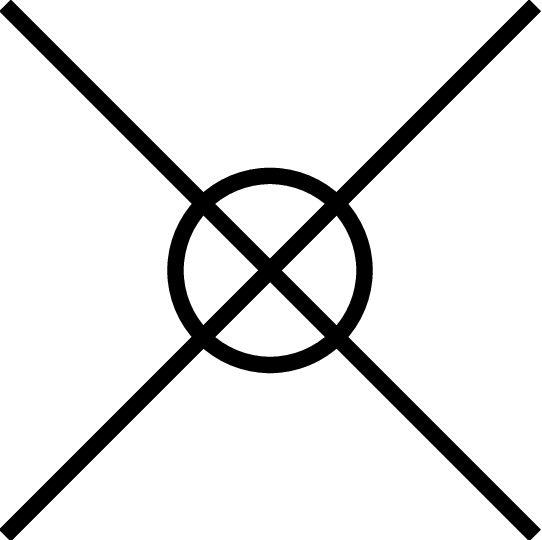}}&=&-i\delta_\lambda.
\end{eqnarray}
To calculate the counterterms $\delta_m$ and $\delta_z$ up to first order in the coupling constant $\lambda$, we start with the two-point function, which is given by:
\begin{eqnarray}\label{Two.point.function}
     \raisebox{-3mm}{\includegraphics[width=1.2cm]{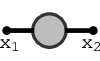}}=\raisebox{-1.5mm}{\includegraphics[width=1.2cm]{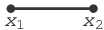}}
   +\raisebox{-2.5mm}{\includegraphics[width=1cm]{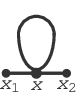}}+\raisebox{-2.5mm}{\includegraphics[width=1.2cm]{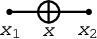}}\hspace{0cm}.
\end{eqnarray}
The standard renormalization condition used to determine the counterterms is typically expressed as:
\begin{eqnarray}\label{Renorm.Condition.}
    \raisebox{-2mm}{\includegraphics[width=1cm]{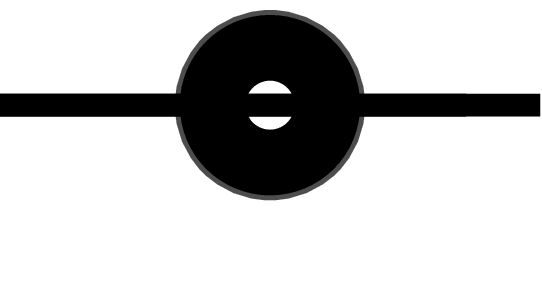}}&=&\frac{i}{p^2-m^2}+\mbox{\tiny(the terms regular at $p^2=m^2$)},\nonumber\\
   \raisebox{-2mm}{\includegraphics[width=0.7cm]{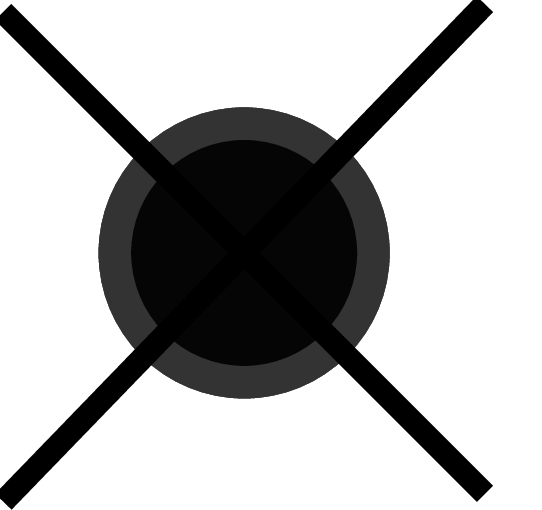}}&=&-i\lambda\hspace{2cm} \mbox{(at $s=4m^2$, $t=u=0$)}.
\end{eqnarray}
The parameters $s$, $t$, and $u$ represent different channels. As is well-known, the type of channel can be identified from the structure of the Feynman diagram, with each channel exhibiting distinct angular dependencies in the cross-section. Using the two-point function given in Eq.~(\ref{Two.point.function}) and the renormalization condition outlined in Eq.~(\ref{Renorm.Condition.}), we determine that the counterterms $\delta_z$ and $\delta_\lambda$ vanish up to first order in the coupling constant $\lambda$. However, the mass counterterm $\delta_m$ is non-zero and is determined as follows:
\begin{eqnarray}\label{mass.counterterm.}
     \delta_m(x)=\frac{-i}{2}\raisebox{-2mm}{\includegraphics[width=1cm]{15.eps}}=\frac{-\lambda}{2}G(x,x),
\end{eqnarray}
where the function $G(x,x')$ denotes the Green's function. The vacuum energy expression is given by:
\begin{eqnarray}\label{Vacuum.En.EXP1.}
     E^{(1)}_{\mbox{\tiny vac.}}&=&i\int d^3\mathbf{x}\bigg(\frac{1}{8} \raisebox{-7mm}{\includegraphics[width=0.5cm]{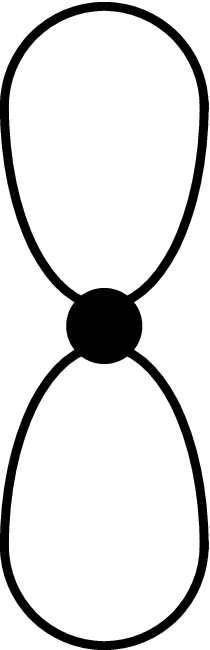}}
       +\frac{1}{2}\raisebox{-1mm}{\includegraphics[width=0.5cm]{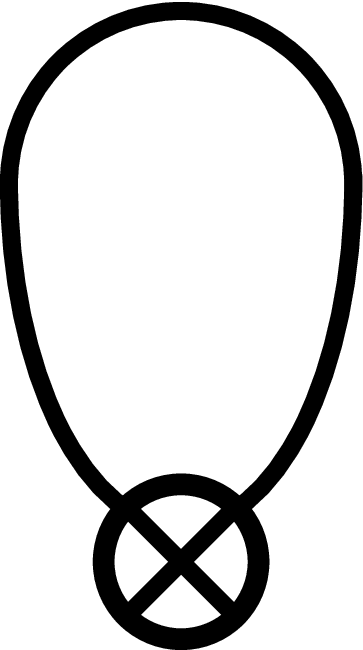}}+...\bigg)
       \\&=&i \int d^3\mathbf{x}\bigg(\frac{-i\lambda}{8}G^2(x,x)-\frac{-i}{2}\delta_m(x)G(x,x)\bigg),\nonumber
\end{eqnarray}
where the superscript $(1)$ on the vacuum energy indicates the first order in the coupling constant $\lambda$. Substituting Eq. (\ref{mass.counterterm.}) into Eq. (\ref{Vacuum.En.EXP1.}) yields:
\begin{eqnarray}\label{Vacuum.En.EXP2.}
  E^{(1)}_{\mbox{\tiny vac.}}=\frac{-\lambda}{8}\int d^3\mathbf{x} G^2(x,x).
\end{eqnarray}

\section{Casimir Energy}\label{Sec.Casimir.En.}
This section details the computation of the leading and next-to-leading orders of the Casimir energy for a massive/massless Lorentz-violating scalar field confined between two parallel rough membranes with four types of boundary conditions~(Dichlet, Neumann, Mixed, and PBC) in $3+1$ dimensions. The following subsection will address the leading order of the Casimir energy, while the subsequent subsection will cover the radiative correction to the Casimir energy.
\subsection{Leading Order}\label{leading.order}
For all cases of boundary conditions~(Dichlet, Neumann, Mixed, and PBC), the vacuum energy of region \(A1\) in Fig. (\ref{fig.BSS}) can be expressed using the following standard form:
\begin{eqnarray}\label{ZO.Vac.Def}
        E_{\mbox{\tiny Vac.$\mathcal{B}$}}^{\mbox{\tiny (0)A1}}=\frac{1}{2\sqrt{1+\sigma_0}}\sum_{n_\mathcal{B}}\int\frac{d^2\mathbf{k}}{(2\pi)^2}\left[p^{(0)}_{n,\mbox{\tiny$\mathcal{B}$}}
        +p^{(1)}_{n,\mbox{\tiny$\mathcal{B}$}}+m^2\right]^{1/2}.
\end{eqnarray}
By applying the definition of the BSS given in Eq. (\ref{BSS.DEF}) and utilizing Eqs. (\ref{eigenvalue.0.D&M}) and (\ref{First.Order.Eigenvalue}), along with changing of variables $\kappa_1=k_1\sqrt{1-\sigma_1}/L$, $\kappa_2=k_2\sqrt{1-\sigma_2}/L$, and $\tilde{a}_0=\frac{a}{C_{\mbox{\tiny$\mathcal{B}$}}\sqrt{1-\sigma_3}\sqrt{1+M}}$, we obtain
\begin{eqnarray}\label{ZO.Cas.1}
     \Delta E_{\mbox{\tiny Vac.,$\mathcal{B}$}}^{\mbox{\tiny (0)}}=E_{\mbox{\tiny Vac.$\mathcal{B}$}}^{\mbox{\tiny(0)}A}-E_{\mbox{\tiny Vac.$\mathcal{B}$}}^{\mbox{\tiny(0)B}}=\frac{L^2}{4\pi\sqrt{1+\sigma_0}\sqrt{1-\sigma_1}\sqrt{1-\sigma_2}}\sum_{n_\mathcal{B}}g_{\mbox{\tiny$\mathcal{B}$}}(n),
\end{eqnarray}
where
\begin{eqnarray}\label{g(n)}
          g_{\mbox{\tiny$\mathcal{B}$}}(n)=\int_{0}^{\infty}\kappa d\kappa\sqrt{\kappa^2+\Big(\frac{k_{n,\mbox{\tiny$\mathcal{B}$}}}{C_{\mbox{\tiny$\mathcal{B}$}}\tilde{a}_0}\Big)^2+m^2}+2\times\{a\to\mbox{\small$\frac{L-a}{2}$}\}-\{a\to b\}-2\times\{a\to\mbox{\small$\frac{L-b}{2}$}\}.
\end{eqnarray}
For the case of PBC, the value of the parameter \( C_{\mathcal{B}} = C_{\mathcal{P}} = 2 \); for all other boundary conditions, the value of \( C_{\mathcal{B}} \) should be considered as 1. For \(\mbox{\small$\mathcal{B}=\{\mathcal{D,N,P}\}$}\), the appropriate form of the Abel-Plana Summation Formula (APSF) that can regularize the summation expression given in Eq.~ (\ref{ZO.Cas.1}) is:
\begin{eqnarray}\label{APSF}
      \sum_{n=1}^{\infty}g_{\mbox{\tiny$\mathcal{B}$}}(n)=\frac{-1}{2}g_{\mbox{\tiny$\mathcal{B}$}}(0)+\int_{0}^{\infty}g_{\mbox{\tiny$\mathcal{B}$}}(x)dx
      +i\int\frac{g_{\mbox{\tiny$\mathcal{B}$}}(it)-g_{\mbox{\tiny$\mathcal{B}$}}(-it)}{e^{2\pi t}-1}dt.
\end{eqnarray}
The appropriate form of the Abel-Plana summation formula for the case of MBC, which effectively regularizes the divergences arising in the summation of $g_{\mbox{\tiny$\mathcal{M}$}}(n+\mbox{$\frac{1}{2}$})$, is given by:
\begin{eqnarray}\label{APSF.MBC}
      \sum_{n=0}^{\infty}g_{\mbox{\tiny$\mathcal{M}$}}(n+\mbox{$\frac{1}{2}$})=\int_{0}^{\infty}g_{\mbox{\tiny$\mathcal{M}$}}(x)dx
      -i\int\frac{g_{\mbox{\tiny$\mathcal{M}$}}(it)-g_{\mbox{\tiny$\mathcal{M}$}}(-it)}{e^{2\pi t}+1}dt,
\end{eqnarray}
In both forms of the APSF, as given in Eqs. (\ref{APSF}) and (\ref{APSF.MBC}), all terms on the right-hand side, except for the last one—referred to as the branchcut term—contain divergent contributions~\cite{APSF.Ref.1,APSF.Ref.2,APSF.Ref.3}. However, after applying the APSF on the summation expression given in Eq. (\ref{ZO.Cas.1}) and implementing the BSS for vacuum energies, all divergent contributions are systematically eliminated~\cite{JHEP.REZA,LP.MAN}. Consequently, only the branchcut terms, which are free from divergences, remain. This process is valid for all types of boundary conditions. As a result, Eq. (\ref{ZO.Cas.1}) simplifies to:
\begin{eqnarray}\label{ZO.Only.Branchcuts.}
          \Delta E_{\mbox{\tiny Vac.,$\mathcal{B}$}}^{\mbox{\tiny (0)}}=\frac{C_{\mbox{\tiny$\mathcal{B}$}}L^2}{4\pi\sqrt{1+\sigma_0}\sqrt{1-\sigma_1}\sqrt{1-\sigma_2}}
          B_{\mbox{\tiny$\mathcal{B}$}}(\tilde{a}_0,m)+2\times\{a\to\mbox{\small$\frac{L-a}{2}$}\}-\{a\to b\}-2\times\{a\to\mbox{\small$\frac{L-b}{2}$}\},
\end{eqnarray}
The branchcut term $B_{\mbox{\tiny$\mathcal{B}$}}(\tilde{a}_0,m)$, after the change of variable $\tau=t\pi/\tilde{a}_0$, transforms into:
\begin{eqnarray}\label{Branchcut.ZO}
       B_{\mbox{\tiny$\mathcal{B}$}}(\tilde{a}_0,m)=\frac{-2c\tilde{a}_0}{\pi}\int_{m}^{\infty}d\tau\int_{0}^{\sqrt{\tau^2-m^2}}\frac{\kappa d\kappa\sqrt{\tau^2-\kappa^2-m^2}}{e^{2\tilde{a}_0\tau}-c},
\end{eqnarray}
wherein $c=\pm1$. The branchcut value for the case of DBC/NBC/PBC corresponds to $c=1$, while $c=-1$ refers to the MBC case. To proceed with the computation of Eq.~(\ref{ZO.Only.Branchcuts.}), the final step involves evaluating the limits embedded in the definition of the BSS, as given in Eq.~(\ref{BSS.DEF}). After computing these limits, the Casimir energy density expression for a massive Lorentz-violating scalar field confined by DBC/MBC between two parallel rough membranes in three spatial dimensions can be expressed as:
\begin{eqnarray}\label{ZO.Casimir.En.Massive}
      \mathcal{E}_{\mbox{\tiny Cas.\mbox{\tiny$\mathcal{B}$}}}^{\mbox{\tiny (0)}}=\frac{-C_{\mbox{\tiny$\mathcal{B}$}}m^2}{8\pi^2\mathcal{A}}
      \sum_{j=1}^{\infty}\frac{c^j K_2(2m\tilde{a}_0j)}{\tilde{a}_0^2j^2},
\end{eqnarray}
where $\mathcal{A}=[(1+\sigma_0)(1-\sigma_1)(1-\sigma_2)(1-\sigma_3)(1+M)]^{1/2}$ and $M=\int_{\frac{-1}{2}}^{\frac{1}{2}}\int_{\frac{-1}{2}}^{\frac{1}{2}}dv_1dv_2\mathcal{M}_a(v_1,v_2)$. For the massless limit of the scalar field, the above expression for the Casimir energy density simplifies to:
\begin{eqnarray}\label{ZO.Casimir.En.Massless}
      \mathcal{E}_{\mbox{\tiny Cas.\mbox{\tiny$\mathcal{B}$}}}^{\mbox{\tiny (0)}}\buildrel {m\to0}\over{\xrightarrow{\hspace{1cm}}}\frac{-C_{\mbox{\tiny$\mathcal{B}$}}\text{Li}_4(c)}{16\pi^2\tilde{a}_0^4\mathcal{A}}=\left\{
      \begin{array}{ll}
        \frac{-C_{\mbox{\tiny$\mathcal{B}$}}\pi^2}{1440\tilde{a}_0^4\mathcal{A}},\hspace{1cm} \hbox{\small for the case of DBC/NBC/PBC;} \\  \\
        \frac{7\pi^2}{11520\tilde{a}_0^4\mathcal{A}}\hspace{1cm}\hbox{\small for the case of MBC,}
      \end{array}\right.
\end{eqnarray}
where $\text{Li}_4(c)$  is the polylogarithm function. The expressions for the Casimir energy density in both the massive and massless cases, as given by Eqs. (\ref{ZO.Casimir.En.Massive}) and (\ref{ZO.Casimir.En.Massless}), reveal that the roughness of the membranes acts similarly to a scale parameter for Lorentz symmetry breaking in the Casimir energy. However, a key distinction between the roughness parameter \(M\) and the Lorentz-violating parameters \(\sigma_i\) is that \(M\) is dependent on the distance between the membranes. It's important to note that in the absence of Lorentz symmetry breaking (\(\sigma_i = 0\)) and with smooth membranes (\(\mathcal{M} = 0\)), our results align with those found in the existing literature (see, for example, \cite{JHEP.REZA,LP.MAN}). Figure (\ref{PLOT.Zero.Deviation}) simultaneously plots the leading-order Casimir energy density for massive and massless scalar fields confined between two rough and smooth membranes under DBC, NBC, PBC, and MBC. This simultaneous plotting allows for a clear graphical representation of the impact that the roughness of the membranes has on the Casimir energy.
\par
In the case of a massless scalar field without Lorentz violation (i.e., \(\sigma_i = 0\)), the difference in the Casimir energy value between the scenarios with and without the roughness properties of the membranes can be expressed as follows:
\begin{eqnarray}\label{Zo.Diff.Cas.per cent}
         \Delta \mathcal{E}_{\mbox{\tiny Cas.$\mathcal{B}$}}^{\mbox{\tiny (0)}}=\frac{-C_{\mbox{\tiny$\mathcal{B}$}}\text{Li}_4(c)}{16\pi^2a^4}[(1+M)^{3/2}-1].
\end{eqnarray}
The ratio of this difference in Casimir energy, relative to the original Casimir energy value in the scenario where there are no roughness properties on the membranes and no Lorentz violations for the scalar field, is given by:
\begin{eqnarray}\label{RATIO.Definition}
\text{Ratio} = \frac{\mathcal{E}_{\text{Cas.}}^{\text{rough}} - \mathcal{E}_{\text{Cas.}}^{\text{smooth}}}{\mathcal{E}_{\text{Cas.}}^{\text{smooth}}}.
\end{eqnarray}
This ratio quantifies the relative impact of the membrane roughness on the Casimir energy when compared to the idealized case of perfectly smooth membranes without any Lorentz symmetry breaking. By expanding this ratio for small values of $M$, we obtain:
\begin{eqnarray}\label{Zo.Diff.Cas.per cent.Ratio}
        \text{Ratio}=\frac{\Delta \mathcal{E}_{\mbox{\tiny Cas.$\mathcal{B}$}}^{\mbox{\tiny (0)}}}{\mathcal{E}_{\mbox{\tiny Cas.$\mathcal{B}$}}^{\mbox{\tiny (0)}}}=(1+M)^{3/2}-1\buildrel{M\to 0}\over {\xrightarrow {\hspace{1.3cm}}}\frac{3}{2}M+\frac{3}{8}M^2+\mathcal{O}(M^3).
\end{eqnarray}
Here, the parameter \(\mbox{\small$\mathcal{B} = \{\mathcal{D}, \mathcal{N}, \mathcal{P}, \mathcal{M}\}$}\) denotes the type of boundary condition applied, where \(\mbox{\small$\mathcal{D}$}\), \(\mbox{\small$\mathcal{N}$}\), \(\mbox{\small$\mathcal{P}$}\), and \(\mbox{\small$\mathcal{M}$}\) correspond to Dirichlet, Neumann, periodic, and mixed boundary conditions, respectively. In the left plot of Fig. (\ref{PLOT.Zero.Ratio}), the relative change in the Casimir energy density, considering both the presence and absence of membrane roughness, is depicted as a function of the membrane separation. In this plot, Lorentz violation for the scalar field was neglected. As demonstrated in Eq. (\ref{Zo.Diff.Cas.per cent.Ratio}), for the massless case, the effect of roughness on the Casimir energy becomes more pronounced as the membrane separation decreases. A similar trend is observed in the massive cases. Furthermore, as depicted in Fig. (\ref{PLOT.Zero.Ratio}), the absolute value of the relative change in Casimir energy density is greater for massive fields than for massless ones. This indicates that surface roughness has a more substantial impact on the Casimir energy of massive scalar fields compared to massless ones. It should be noted that, since the maximum value of the function \(h(x,y)\) must be much smaller than the membrane separation (\emph{i.e.}, \( \text{Max}\{h(x,y)\} \ll a \)), the validity of the plots in regions where \( a \leq h(x,y) \) cannot be trusted. Consequently, these regions are indicated by dashed lines in the plots.
\subsection{Radiative Correction}\label{SubSec:RC}
\begin{figure}[th]\centering\includegraphics[width=8cm]{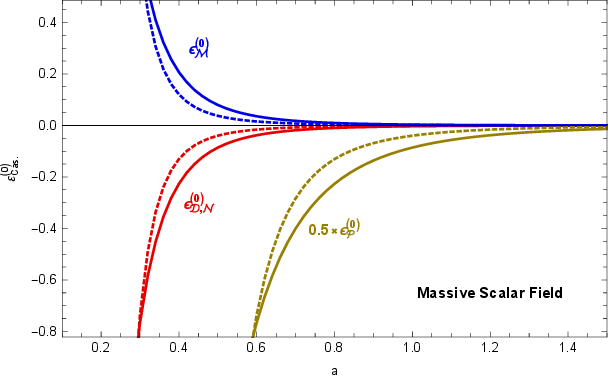}\includegraphics[width=8cm]{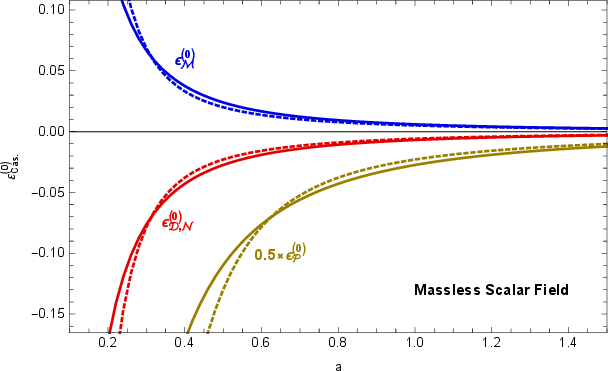}
\caption{\label{PLOT.Zero.Deviation}
The left (right) figure depicts the leading-order Casimir energy density for a massive (massless) scalar field confined between two membranes under Dirichlet, Neumann, periodic, and mixed boundary conditions, with one membrane being smooth and the other rough. The energy is plotted as a function of the membrane separation distance \(a\) in three spatial dimensions. The solid lines represent the scenario where the membranes are smooth, and neither roughness nor Lorentz violation is present, while the dashed lines correspond to the case where the scalar field experiences Lorentz violation and the membranes exhibit roughness. The Lorentz violation factor in these plots is \(\sigma_{i=0,1,2,3} = 0.1\). In the left figure, the mass of the scalar field is set to \(m = 1\). The roughness of the membranes is modeled by the function \(h(x,y) = \frac{1}{4}\cos(x\pi/2)\cos(y\pi/2)\). These plots demonstrate that the deviation in Casimir energy due to the presence of Lorentz violation and membrane roughness is substantial, reaching approximately 40\% of the Casimir energy value in certain regions for both massive and massless scalar fields. The units for all parameters in the plots are chosen such that $\hbar c=1$. }
\end{figure}

\begin{figure}[th]\centering\includegraphics[width=8cm]{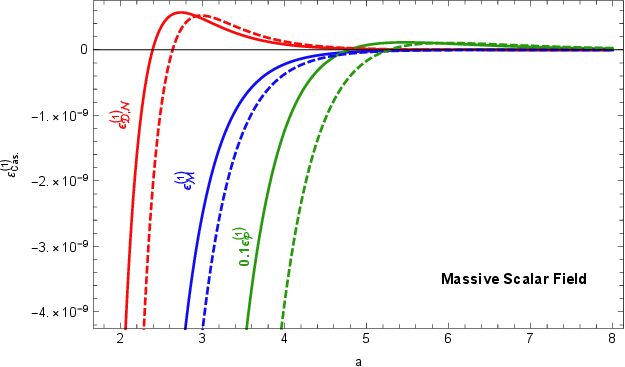}\includegraphics[width=8cm]{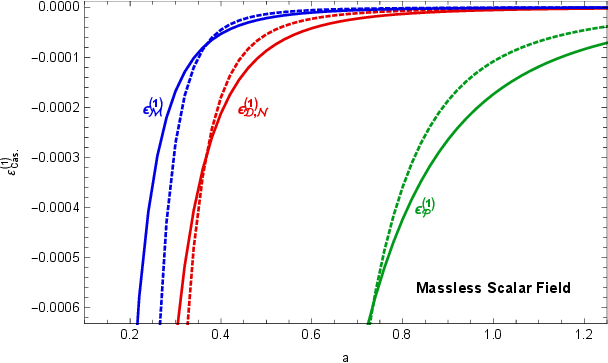}
\caption{\label{PLOT.RC.Deviation}
The left (right) figure illustrates the radiative correction to the Casimir energy density for a massive (massless) scalar field confined between two membranes under Dirichlet, Neumann, periodic, and mixed boundary conditions, with one membrane smooth and the other rough. The energy is plotted as a function of the membrane separation distance \(a\) in three spatial dimensions. The solid lines represent the scenario with smooth membranes, without roughness or Lorentz violation, while the dashed lines correspond to the case where the scalar field experiences Lorentz violation and the membranes are rough. In these plots, the Lorentz violation factors are set to \(\sigma_{i=0,1,2,3} = 0.1\), the coupling constant is \(\lambda = 0.1\), and the mass of the scalar field in the left figure is \(m = 1\). The membrane roughness is modeled by the function \(h(x,y) = \frac{1}{4}\cos(x\pi/2)\cos(y\pi/2)\). These plots indicate that the deviation in the Casimir energy due to the combined effects of Lorentz violation and membrane roughness is significant. The units for all parameters in the plots are set by assuming $\hbar c=1$.}
\end{figure}

\begin{figure}[th]\centering\includegraphics[width=8cm]{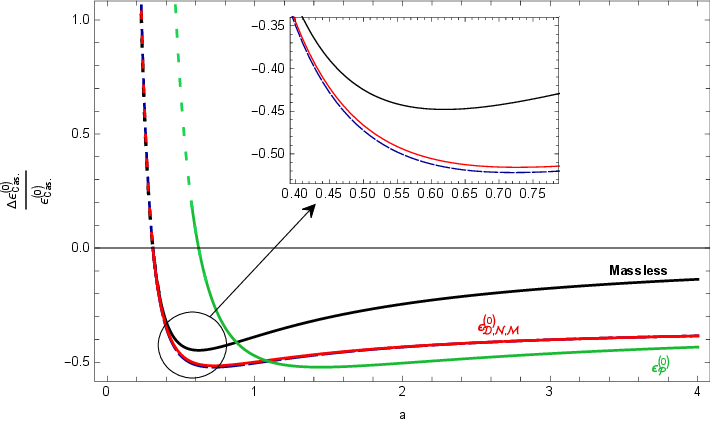}\includegraphics[width=8cm]{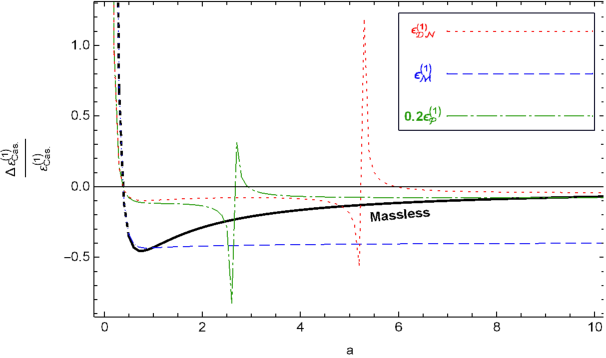}
\caption{\label{PLOT.Zero.Ratio}
The left (right) figure presents the ratio of changes in the leading-order Casimir energy density (radiative correction to the Casimir energy) for a scalar field confined between two parallel membranes, with and without roughness, relative to the Casimir energy with smooth membranes, plotted as a function of the membrane separation distance \(a\). These plots display results for all boundary conditions (DBC, NBC, PBC, and MBC) for both massive and massless scalar fields. The solid line (black line) in both figures corresponds to the massless cases. In the right figure, due to the sign change in the radiative correction, singularities appear in regions where the Casimir energy changes sign, indicating that the impact of membrane roughness in these regions is particularly significant. Lorentz violation is turned off in all plots (\(\sigma_{i=0,1,2,3} = 0\)). The coupling constant is set to \(\lambda = 0.1\), and the mass of the field is \(m = 1\). The roughness of the membranes is modeled by the function \(h(x,y) = \frac{1}{4}\cos(x\pi/2)\cos(y\pi/2)\). The units for all parameters in the plots are based on the convention $\hbar c=1$.}
\end{figure}

First, it is important to recall that the radiative correction to the Casimir energy expression for a massive/massless scalar field, without Lorentz violation (i.e., $\sigma_{i=\{0,1,2,3\}} = 0$), confined by DBC/NBC/MBC/PBC between two smooth parallel plates (i.e., $\mathcal{M}_a(v_1,v_2) = 0$) with separation $a$ in three spatial dimensions, has been previously reported in the literature~\cite{JHEP.REZA,mixed3D.MAN,LP.MAN}. In the reported work, the renormalization procedure employed relies on position-dependent counterterms, and the vacuum energy was extracted using these counterterms. Additionally, the authors applied the BSS in conjunction with the cutoff regularization technique to address the divergences that arise during Casimir energy calculations. A brief overview of these computational steps is provided in Appendix \ref{Appendix.RC.Calculation}. The result for the radiative correction to the Casimir energy density in the case of a massive scalar field is:
\begin{eqnarray}\label{general.form.DBC&MBC.RC.}
             \mathcal{E}_{\mbox{\tiny Cas.$\mathcal{B}$}}^{\mbox{\tiny (1)}}(m,a)=\frac{-\lambda C_{\mbox{\tiny$\mathcal{B}$}} m}{128\pi^3a}\sum_{j=1}^{\infty}\frac{c^j K_1(2maj/C_{\mbox{\tiny$\mathcal{B}$}})}{j}\left[\frac{m C_{\mbox{\tiny$\mathcal{B}$}}}{\pi a}\sum_{j'=1}^{\infty}\frac{c^{j'} K_1(2maj'/C_{\mbox{\tiny$\mathcal{B}$}})}{j'}+(\mbox{\large$\frac{c+1}{2}$})\frac{C_{\mbox{\tiny$\mathcal{B}$}}m}{a}-\frac{m^2}{\pi}(\mbox{$\frac{1}{2}$}+\ln2)\right],
\end{eqnarray}
where \(\mbox{\small$\mathcal{B}$}=\{\mbox{\small$\mathcal{D}$},\mbox{\small$\mathcal{N}$},\mbox{\small$\mathcal{P}$},\mbox{\small$\mathcal{M}$}\}\), $c = \pm 1$, and $m$ denotes the mass of the scalar field. For a scalar field confined with DBC/NBC/PBC, we set $c = 1$, while $c = -1$ corresponds to the case with MBC. Furthermore, the parameter \(C_{\mbox{\tiny$\mathcal{B}$=\{$\mathcal{D}$,$\mathcal{N}$,$\mathcal{P}$,$\mathcal{M}$\}}}=\{1,1,2,1\}\). For the massless scalar field, the reported result is:
\begin{eqnarray}\label{ECas.RC.Massless Without LV&Roughness}
        \mathcal{E}_{\mbox{\tiny Cas.$\mathcal{B}$}}^{\mbox{\tiny (1)}}(0,a)=\frac{-\lambda C^4_{\mbox{\tiny$\mathcal{B}$}} \text{Li}_2(c)^2}{512\pi^4a^4}=\left\{
         \begin{array}{ll}
           \frac{-\lambda C^4_{\mbox{\tiny$\mathcal{B}$}} }{18432a^4},\hspace{1cm}\hbox{\small for the case of DBC/NBC/PBC;} \\
           \hspace{1cm} \\
           \frac{-\lambda }{73728a^4},\hspace{1cm}\hbox{\small for the case of MBC.}
         \end{array}\right.
\end{eqnarray}
Our goal in this section is to determine the radiative correction to the Casimir energy for a massive/massless Lorentz-violating scalar field confined by DBC/NBC/PBC/MBC between two rough parallel membranes in $3+1$ dimensions. To achieve this, we start by using the vacuum energy subtraction method described in Eq. (\ref{BSS.DEF}). By substituting the Green's function expression given in Eq. (\ref{Greens.function.DBC.LV&Roughness}) into Eq. (\ref{Vacuum.En.EXP2.}), we obtain:
\begin{eqnarray}\label{RC.Vac.En.1}
      \Delta \tilde{E}_{\mbox{\tiny Vac.$\mathcal{B}$}}^{(1)}=\frac{-\lambda}{8(1+\sigma_0)(1-\sigma_1)(1-\sigma_2)}\int G_{\mbox{\tiny$\mathcal{B}$}}(\tilde{a}_1;x,x')^2d^3\mathbf{x}+2E^{(1)}_{\mbox{\tiny Vac.A2}}-E^{(1)}_{\mbox{\tiny Vac.B1}}-2E^{(1)}_{\mbox{\tiny Vac.B2}}.
\end{eqnarray}
The first term on the right-hand side of Eq. (\ref{RC.Vac.En.1}) represents the vacuum energy of region \(A1\) in Fig. (\ref{fig.BSS}), now expressed using the updated Green's function given in Eq. (\ref{Greens.function.DBC.LV&Roughness}). This expression incorporates both the roughness of the membranes and the Lorentz violation of the scalar field. From this point forward, the calculation of the radiative correction to the Casimir energy follows a similar approach to that outlined in Appendix B. Therefore, using the results derived in that appendix, the radiative correction to the Casimir energy for a massive scalar field confined between two rough membranes can be expressed as follows:
\begin{eqnarray}\label{Final.ECas.RC.Roughness&LV}
        \tilde{\mathcal{E}}_{\mbox{\tiny Cas.$\mathcal{B}$}}^{(1)}(a,m)=\frac{\mathcal{E}_{\mbox{\tiny Cas.$\mathcal{B}$}}^{(1)}(\tilde{a}_1,m)}{(1+\sigma_0)(1-\sigma_1)(1-\sigma_2)},
\end{eqnarray}
where $\mathcal{E}_{\mbox{\tiny Cas.}}^{(1)}(\tilde{a}_1,m)$ denotes the radiative correction term of the Casimir energy in the absence of Lorentz violation and boundary roughness, as given in Eq. (\ref{general.form.DBC&MBC.RC.}). Furthermore, $\tilde{a}_1=\frac{a}{\sqrt{1-\sigma_3}\sqrt{1+\mathcal{M}_a(0,0)}}$. It is important to note that the relation provided by Eq. (\ref{Final.ECas.RC.Roughness&LV}) is also applicable to the radiative correction of the Casimir energy for the massless cases. %\par Given the known differences in the wave number values between NBC/PBC and DBC, the Casimir energy density for NBC/PBC can be expressed in terms of the Casimir energy density obtained for DBC as follows:
%\begin{eqnarray}\label{Final.ECas.RC.NBC&MBC}
%      \tilde{\mathcal{E}}_{\mbox{\tiny Cas.$\mathcal{N}$}}^{(1)}(a,m)&=&\tilde{\mathcal{E}}_{\mbox{\tiny Cas.$\mathcal{D}$}}^{(1)}(a,m), \nonumber\\
%      \tilde{\mathcal{E}}_{\mbox{\tiny Cas.$\mathcal{P}$}}^{(1)}(a,m)&=&2\tilde{\mathcal{E}}_{\mbox{\tiny Cas.$\mathcal{D}$}}^{(1)}(\mbox{\small $\frac{a}{2}$},m).
%\end{eqnarray}
It should be emphasized that, the relation discussed above is equally applicable to the zero-order Casimir energy. In Fig.~(\ref{PLOT.RC.Deviation}), we present plots of the radiative correction to the Casimir energy density as a function of membrane separation distance $a$, comparing scenarios both with and without membrane roughness and Lorentz violation of the scalar field. The solid lines represent the Casimir energy where both Lorentz violation and membrane roughness are omitted, while the dashed lines account for the effects of both roughness and Lorentz violation. In the left diagram, the radiative corrections to the Casimir energy for the massive scalar field are compared, while the right diagram illustrates the results for the massless case. As shown in the plots in Figs. (\ref{PLOT.Zero.Deviation}) and (\ref{PLOT.RC.Deviation}), membrane roughness has a substantial impact on the Casimir energy across all four types of boundary conditions, influencing both the zero-order term and the radiative corrections. Notably, even when the roughness function $h(x,y)$ is assigned a small value, the impact on the Casimir energy remains considerable due to the roughness properties of the membranes. To better understand the extent of deviation in the zero- and first-order radiative corrections to the Casimir energy density caused solely by membrane roughness, the relative change in Casimir energy was plotted in the left and right panels of Fig.~(\ref{PLOT.Zero.Ratio}). All plots in Fig.~(\ref{PLOT.Zero.Ratio}) were generated with Lorentz violation effects omitted. For the cases of Dirichlet, Neumann, and periodic boundary conditions (DBC, NBC, and PBC), the radiative correction to the Casimir energy changes sign. The right panel of Fig.~(\ref{PLOT.Zero.Ratio}) demonstrates that in the region where the Casimir energy changes sign, the relative changes in the Casimir energy become particularly noticeable. In the massless case, as shown in Fig.~(\ref{PLOT.Zero.Ratio}), as the membrane separation decreases, the deviation due to roughness can reach up to approximately 40\% of the Casimir energy value observed in the absence of roughness.
\par
It is crucial to note that the maximum value of the function $h(x,y)$ must be much smaller than the membrane separation ($Max\{h(x,y)\}\ll a$). As such, the validity of the plots in regions where $a\leq h(x,y)$ cannot be guaranteed. These regions are therefore indicated by dashed lines in the plots to signal caution.

\section{Conclusions}\label{Sec.Conclusion}
In this paper, we computed the zero- and first-order radiative corrections to the Casimir energy for a self-interacting massive/massless scalar field confined between two rough membranes under Dirichlet,Neumann, Periodic and mixed boundary conditions. Typically, two distinct approaches are presented in the literature for calculating the radiative correction to the Casimir energy of a massive or massless scalar field confined between two smooth parallel membranes. The first approach employs a renormalization program that uses free-space counterterms to renormalize the bare parameters of the Lagrangian~\cite{Toms,MBCruz,free.counterterms.1,free.counterterms.2,free.counterterms.3,free.counterterms.4}. In contrast, the second approach utilizes position-dependent counterterms to incorporate the effects of boundary conditions into the renormalization procedure~\cite{1D.Reza,EUR-Reza}. In the second approach, a systematic framework is applied to renormalize the bare parameters of the Lagrangian, effectively accounting for boundary influences. In this work, we adopted the second renormalization program, combined with the BSS as a regularization technique, to provide a clear and unambiguous method for calculating the Casimir energy. Consequently, our findings for the radiative correction to the Casimir energy, after incorporating the effects of membrane roughness and Lorentz violation, differ from those obtained using the first approach. However, they are consistent with the results reported in the second series of studies~\cite{Ddim.MAN,mixed3D.MAN}, which also employed position-dependent counterterms. Comparing these results to the case of smooth membranes without roughness reveals that roughness can significantly alter the Casimir energy at both the zero-order and radiative correction levels. This deviation in the Casimir energy increases as the distance between the membranes decreases, with the maximum deviation in the zero- and first-order corrections approaching approximately 40\% of the Casimir energy for smooth membranes.
\appendix
\section{Calculation of Green's Function in Presence of Rough Membrames}\label{Appendix.GreensFunc.}
\setcounter{equation}{0}
In this appendix, we provide a detailed computation of the Green's function for a massive Lorentz-violating scalar field confined between two parallel rough membranes in \(3+1\) dimensions under Dirichlet, Neumann, Periodic and MBCs. To achieve this, we start by considering Eq. (\ref{equation.O.motion}) and proceed to obtain
\begin{eqnarray}\label{Equation.1Green}
        \Big[(1+\sigma_0)\partial_0^2-\mathbf{P}+m^2\Big]G_{\mbox{\tiny$\mathcal{B}$}}(a;v,v')=\frac{1}{L^2a}\delta(v_1-v'_1)\delta(v_2-v'_2)\delta(v_3-v'_3)
\end{eqnarray}
where
\begin{eqnarray}\label{1st.Form.Green}
        G_{\mbox{\tiny$\mathcal{B}$}}(a;v,v')=\sum_{n_{\mbox{\tiny$\mathcal{B}$}}}
        \int\frac{d\omega}{2\pi}\int\frac{d^2\mathbf{k}}{(2\pi)^2}\mathcal{C}_{n,\mbox{\tiny$\mathcal{B}$}}(v')\phi^{(0)}_{n,\mbox{\tiny$\mathcal{B}$}}(v),
\end{eqnarray}
where $v=(v_1,v_2,v_3)$ and $v'=({v'}_1,{v'}_2,{v'}_3)$. Furthermore, for DBC/NBC, the domain of $n_\mathcal{B}$ is $n_{\mbox{\tiny$\mathcal{D,N}$}}\in\mathbb{N}$, while for the case of MBC, the domain is $n_\mathcal{M}\in\mathbb{N}\cup\{0\}$. This allowed domain for the case of PBC is $n_{\mbox{\tiny$\mathcal{P}$}}\in\mathbb{Z}-\{0\}$. To determine the form of the function \(\mathcal{M}_a(v_1,v_2)\) embedded in the operator \(\mathbf{P}\), we select the function \(h(v_1,v_2)\), which characterizes the roughness properties of the membranes, as follows:
\begin{eqnarray}\label{M.function}
         h(v_1,v_2)=\epsilon_1\epsilon_2\cos(\alpha_1Lv_1+\theta_1)\cos(\alpha_2Lv_2+\theta_2)
\end{eqnarray}
Here, \(\epsilon_1\) and \(\epsilon_2\) denote the roughness domains on the membranes, and their values are much smaller than the membrane separation distance \(a\). The parameters \(\alpha_i\) and \(\theta_i\) represent the frequency of roughness and the phase shift, respectively. To determine the coefficient \(\mathcal{C}_{n,\mbox{\tiny$\mathcal{B}$}}(v')\) in Eq. (\ref{1st.Form.Green}), we multiply the wave function \(\phi^{(0)\ast}_{n,\mbox{\tiny$\mathcal{B}$}}(v')\) on the left side of Eq. (\ref{Equation.1Green}). Then, by substituting Eq. (\ref{M.function}) for the function \(\mathcal{M}_a(v_1,v_2)\), we integrate over all spatial directions. Consequently, we obtain
\begin{eqnarray}\label{Equation.2Green}
      \left[-(1+\sigma_0)\omega^2+\frac{(1-\sigma_1){k'}_1^2+(1-\sigma_2){k'}_2^2}{L^2}+(1-\sigma_3)
      \Big(\frac{k_{n',\mbox{\tiny$\mathcal{B}$}}}{a}\Big)^2\left(1+\mathcal{M}_a(0,0)\right)+m^2\right]\mathcal{C}_{n',\mathcal{B}}(v')
      =\frac{1}{L^2a}\phi^{(0)\ast}_{n',\mbox{\tiny$\mathcal{B}$}}(v')
\end{eqnarray}
where
\begin{eqnarray}\label{M00}
\mathcal{M}_a(0,0)&=&\int_{\frac{-1}{2}}^{\frac{1}{2}}dv_1\int_{\frac{-1}{2}}^{\frac{1}{2}}dv_2\int\frac{d^2\mathbf{k}}{(2\pi)^2}
      \mathcal{M}_a(v_1,v_2)e^{i(\mathbf{k}-\mathbf{k}')\cdot \mathbf{v}}=\frac{-2h(0,0)}{a}+\frac{3h^2(0,0)}{a^2}+...\nonumber\\&&=\frac{-2}{a}\epsilon_1\epsilon_2\cos\theta_1\cos\theta_2+\frac{3}{a^2}\epsilon_1^2\epsilon_2^2\cos^2\theta_1\cos^2\theta_2+...
\end{eqnarray}
Substituting the result for the coefficient \(\mathcal{C}_{n,\mbox{\tiny$\mathcal{B}$}}(v)\) from Eq. (\ref{Equation.2Green}) into the Green's function expression given in Eq. (\ref{1st.Form.Green}) yields the form presented in Eq. (\ref{Greens.function.DBC.LV&Roughness}).

\section{Calculation of Radiative Correction}\label{Appendix.RC.Calculation}
In this appendix, we provide detailed calculations of the radiative correction to the Casimir energy for a massive scalar field confined between two parallel plates under Dirichlet, Neumann, Periodic and mixed boundary conditions, employing the BSS as a regularization technique. These calculations were extensively detailed in our previous work \cite{JHEP.REZA,mixed3D.MAN}. Since we build upon the results from that study, we offer a brief overview of the relevant computations here. Starting with the only three of boundary condition \(\mbox{\small$\mathcal{B}$}=\{\mbox{\small$\mathcal{D}$},\mbox{\small$\mathcal{N}$},\mbox{\small$\mathcal{P}$}\}\) and using Eqs. (\ref{BSS.DEF}, \ref{Greens.function.withoutLV&Roughness}, \ref{Vacuum.En.EXP2.}), we obtain:
\begin{eqnarray}\label{Subtraction.Vac.1}
     \Delta E_{\mbox{\tiny Vac.$\mathcal{B}$}}^{(1)}&=&\frac{-\lambda L^2a}{8(2\pi)^6}
     \int_{\frac{-1}{2}}^{\frac{1}{2}}dv_3\frac{4}{a^2}\sum_{n_{\mbox{\tiny$\mathcal{B}$}},n'_{\mbox{\tiny$\mathcal{B}$}}}\int_{0}^{\infty}\frac{4\pi k^2dk}{k^2+\Omega_{n,\mbox{\tiny$\mathcal{B}$}}^2} \int_{0}^{\infty}\frac{4\pi {k'}^2dk'}{{k'}^2+\Omega_{n',\mbox{\tiny$\mathcal{B}$}}^2}
     \left[\mathcal{H}_{n,\mbox{\tiny$\mathcal{B}$}}(v_3)\mathcal{H}_{n',\mbox{\tiny$\mathcal{B}$}}(v_3)\right]^2\nonumber\\&&+2\times\{a\to\mbox{\small$\frac{L-a}{2}$}\}-\{a\to b\}-2\times\{a\to\mbox{\small$\frac{L-b}{2}$}\}
\end{eqnarray}
where $\Omega_{n,\mbox{\tiny$\mathcal{B}$}}^2=\left(\frac{k_{n,\mbox{\tiny$\mathcal{B}$}}}{a}\right)^2+m^2$ and $k=(\omega,\mathbf{k})$. All integrations over \(k\) and \(k'\) are divergent due to the upper limits of the integrals. To regularize this divergence, we introduce a cutoff by replacing the upper limit of the integrations with a finite value. This cutoff is applied consistently across all corresponding integral expressions. By performing the integration over \(k\) and expanding the result in the limit as the cutoff approaches infinity, we obtain:
\begin{eqnarray}\label{cutoff.Reg.}
       \int_{0}^{\Lambda}\frac{k^2dk}{k^2+\Omega^2}=\Lambda -\Omega  \tan ^{-1}\left(\frac{\Lambda }{\Omega }\right)\buildrel {\Lambda\to\infty}\over{\xrightarrow{\hspace{1.2cm}}}\Lambda-\frac{\pi}{2}\Omega+\mathcal{O}(\Lambda^{-1})
\end{eqnarray}
The first term on the right-hand side of the above equation exhibits linear divergence for the integral over \(k\). By appropriately adjusting the cutoff value in the upper limit of integration for the vacuum energy expressions associated with each region depicted in Fig. (\ref{fig.BSS}), we can eliminate the divergent contributions from this term~\cite{LP.MAN}. Consequently, the remaining terms, which are associated with the second term on the right-hand side of Eq. (\ref{cutoff.Reg.}), persist. Thus, we have:
\begin{eqnarray}\label{Subtraction.Vac.2}
      \Delta E_{\mbox{\tiny Vac.$\mathcal{B}$}}^{(1)}=\frac{-\lambda L^2}{128\pi^2a}
     \sum_{n_{\mbox{\tiny$\mathcal{B}$}},n'_{\mbox{\tiny$\mathcal{B}$}}}\Big(1+\frac{1}{2}\delta_{n,n'}\Big)\Omega_{n,\mbox{\tiny$\mathcal{B}$}}\Omega_{n',\mbox{\tiny$\mathcal{B}$}}+2\times\{a\to\mbox{\small$\frac{L-a}{2}$}\}-\{a\to b\}-2\times\{a\to\mbox{\small$\frac{L-b}{2}$}\}
\end{eqnarray}
The summation over \(n\) and \(n'\) causes Eq. (\ref{Subtraction.Vac.2}) to diverge. To regularize these divergences, we convert the summations into integrals using the APSF introduced by Eq. (\ref{APSF}). Thus, we obtain:
\begin{eqnarray}\label{Subtraction.Vac.3}
      \Delta E_{\mbox{\tiny Vac.$\mathcal{B}$}}^{(1)}=\frac{-\lambda C^2_{\mbox{\tiny$\mathcal{B}$}} L^2}{128\pi^2a}&&\hspace{-0.3cm}\left[\left(\frac{-m}{2}+\int_{0}^{\infty}dx\sqrt{(\mbox{\small$\frac{C_{\mbox{\tiny$\mathcal{B}$}}x\pi}{a}$})^2+m^2}+B_1(a,m)\right)^2
      -\frac{m^2}{4}+\frac{1}{2}\int_{0}^{\infty}\big((\mbox{\small$\frac{C_{\mbox{\tiny$\mathcal{B}$}}x\pi}{a}$})^2+m^2\big)dx
      \right.\nonumber\\&&\left.+\frac{1}{2}B_{2,\mbox{\tiny$\mathcal{B}$}}(a,m)\right]
     +2\times\{a\to\mbox{\small$\frac{L-a}{2}$}\}-\{a\to b\}-2\times\{a\to\mbox{\small$\frac{L-b}{2}$}\}
\end{eqnarray}
where \(B_1(a,m)\) and \(B_2(a,m)\) represent the branch cut terms of the APSF (\emph{i.e.}, the last term in Eq. (\ref{APSF})). The value of the branch cut term \(B_2(a,m)\) is zero. However, for the branch cut term \(B_1(a,m)\), we obtain:
\begin{eqnarray}\label{branchcut-1}
       B_1(a,m)=\frac{-2m^2a}{\pi C_{\mbox{\tiny$\mathcal{B}$}}}\int_{1}^{\infty}\frac{(\eta^2-1)^{\frac{1}{2}}}
       {e^{\frac{2ma\eta}{C_{\mbox{\tiny$\mathcal{B}$}}}}-1}d\eta
               =\frac{-m}{\pi}\sum_{j=1}^{\infty}\frac{K_1(2maj/C_{\mbox{\tiny$\mathcal{B}$}})}{j},
\end{eqnarray}
where \(\frac{1}{e^x - 1} = \sum_{j=1}^{\infty} e^{-jx}\) has been used, along with the change of variable \(\eta = t\pi/a\). Expanding the first term in the bracket of Eq. (\ref{Subtraction.Vac.3}) yields:
\begin{eqnarray}\label{Subtraction.Vac.4}
      \Delta E_{\mbox{\tiny Vac.$\mathcal{B}$}}^{(1)}&=&\frac{-\lambda C_{\mbox{\tiny$\mathcal{B}$}} L^2}{128\pi^2}\left[\frac{m^4a}{C_{\mbox{\tiny$\mathcal{B}$}}\pi^2}\left(\int_{0}^{\infty}d\xi\sqrt{\xi^2+1}\right)^2-\frac{m^3}{2\pi}\int_{0}^{\infty}d\xi\big[2\sqrt{\xi^2+1}-\xi^2-1\big]
        +\frac{2m^2B_1(a,m)}{\pi}\int_{0}^{\infty}d\xi\sqrt{\xi^2+1}\right.\nonumber\\&&\left.
      +\frac{C_{\mbox{\tiny$\mathcal{B}$}}B_1(a,m)}{a}[B_1(a,m)-m]+\frac{C_{\mbox{\tiny$\mathcal{B}$}}}{2a}B_2(a,m)\right]
    +2\times\{a\to\mbox{\small$\frac{L-a}{2}$}\}-\{a\to b\}-2\times\{a\to\mbox{\small$\frac{L-b}{2}$}\}
\end{eqnarray}
The first term in the bracket of Eq. (\ref{Subtraction.Vac.4}) is divergent, and its contribution will be removed using the BSS as follows:
\begin{eqnarray}\label{remove.BSS.1}
\left[a+2\frac{L-a}{2}-b-2\frac{L-b}{2}\right]\frac{m^4}{C_{\mbox{\tiny$\mathcal{B}$}}\pi^2}\left(\int_{0}^{\infty}d\xi\sqrt{\xi^2+1}\right)^2=0.
\end{eqnarray}
The second term in the bracket of Eq. (\ref{Subtraction.Vac.4}) is independent of \(a\) and \(b\), and thus its contribution will be automatically removed by the subtraction process provided by the BSS. To regularize the infinities arising from the third term of Eq. (\ref{Subtraction.Vac.4}), we use the cutoff regularization technique. Specifically, we replace the upper limit of the integral in this term, as well as in analogous integrals associated with the vacuum energy of other regions, with a distinct cutoff value. For this purpose, we substitute the upper limit of the integral in the third term of the equation, which corresponds to region \(A1\) in Fig. (\ref{fig.BSS}), with \(\Lambda_{\mbox{\tiny$A1$}}\). Similarly, cutoff values \(\Lambda_{\mbox{\tiny$A2$}}\), \(\Lambda_{\mbox{\tiny$B1$}}\), and \(\Lambda_{\mbox{\tiny$B2$}}\) are used for the integrals associated with regions \(A2\), \(B1\), and \(B2\), respectively. By computing the integral after substituting these cutoff values and expanding the result in the limit as the cutoff approaches infinity, we isolate the divergent parts. For example, we demonstrate this process for the integral term of region \(A1\) as follows:
\begin{eqnarray}\label{remove.BSS.2}
\frac{2m^2B_1(a,m)}{\pi}\int_{0}^{\Lambda_{\mbox{\tiny$A1$}}}d\xi\sqrt{\xi^2+1}&=&\frac{2m^2B_1(a,m)}{\pi}\left[\frac{1}{2} \left(\Lambda_{\mbox{\tiny$A1$}}  \sqrt{\Lambda_{\mbox{\tiny$A1$}} ^2+1}+\sinh ^{-1}(\Lambda_{\mbox{\tiny$A1$}} )\right)\right]\nonumber\\&&\buildrel\Lambda_{\mbox{\tiny$A1$}}\to\infty\over{\xrightarrow{\hspace{1.2cm}}}\frac{2m^2B_1(a,m)}{\pi}\left[\frac{\Lambda_{\mbox{\tiny$A1$}} ^2}{2}+\frac{1}{4}\big(2\ln\Lambda_{\mbox{\tiny$A1$}} +1+\ln4\big)+\mathcal{O}(\Lambda_{\mbox{\tiny$A1$}}^{-2})\right].
\end{eqnarray}
The proper adjustment of the cutoffs for each region ensures that all divergent contributions from the third term of Eq. (\ref{Subtraction.Vac.4}) are removed. The relationship for this adjustment is:
\begin{eqnarray}\label{adjusting.cutoff.DBC.RC.}
\frac{\Lambda_{\mbox{\tiny$A1$}}^2+\ln\Lambda_{\mbox{\tiny$A1$}}}{\Lambda_{\mbox{\tiny$B1$}}^2+\ln\Lambda_{\mbox{\tiny$B1$}}}=\frac{B_1(b,m)}{B_1(a,m)},\hspace{1.2cm}
\frac{\Lambda_{\mbox{\tiny$A2$}}^2+\ln\Lambda_{\mbox{\tiny$A2$}}}{\Lambda_{\mbox{\tiny$B2$}}^2+\ln\Lambda_{\mbox{\tiny$B2$}}}=\frac{B_1(\mbox{\small$\frac{L-b}{2}$},m)}{B_1(\mbox{\small$\frac{L-a}{2}$},m)},
\end{eqnarray}
As a result, the finite part remaining from the third term can be expressed as:
\begin{eqnarray}\label{remove.BSS.22}
\frac{2m^2B_1(a,m)}{\pi}\int_{0}^{\Lambda_{\mbox{\tiny$A1$}}}d\xi\sqrt{\xi^2+1}\buildrel{\mbox{\small finite part}}\over{\xrightarrow{\hspace{1.6cm}}}\frac{m^2B_1(a,m)}{\pi}\big(\mbox{$\frac{1}{2}$}+\ln2\big).
\end{eqnarray}
Applying the finite part obtained from Eq. (\ref{remove.BSS.22}) to Eq. (\ref{Subtraction.Vac.4}) results in:
\begin{eqnarray}\label{after.cancel.DBC.MBC.Appendix}
      \Delta E_{\mbox{\tiny Vac.$\mathcal{B}$}}^{(1)}&=&\frac{-\lambda C_{\mbox{\tiny$\mathcal{B}$}} L^2}{128\pi^2}\left[\frac{C_{\mbox{\tiny$\mathcal{B}$}}B_1(a,m)}{a}[B_1(a,m)-m]+\frac{m^2B_1(a,m)}{\pi}\big(\mbox{$\frac{1}{2}$}+\ln2\big)\right]
     \nonumber\\&&+2\times\{a\to\mbox{\small$\frac{L-a}{2}$}\}-\{a\to b\}-2\times\{a\to\mbox{\small$\frac{L-b}{2}$}\}.
\end{eqnarray}
The final step involves evaluating the limits specified in Eq. (\ref{BSS.DEF}). Upon completing this calculation, the final result for the radiative correction to the Casimir energy of a massive self-interacting scalar field confined by DBC between two parallel plates is obtained as:
\begin{eqnarray}\label{ECas.RC.Massive Without LV&Roughness}
        \mathcal{E}_{\mbox{\tiny Cas.$\mathcal{B}$}}^{\mbox{\tiny (1)}}(m,a)=\frac{-\lambda C_{\mbox{\tiny$\mathcal{B}$}} L^2m}{128\pi^3}\sum_{j=1}^{\infty}\frac{K_1(2maj/C_{\mbox{\tiny$\mathcal{B}$}})}{j}
        \left[\frac{C_{\mbox{\tiny$\mathcal{B}$}}m}{\pi a}\sum_{j'=1}^{\infty}\frac{K_1(2maj'/C_{\mbox{\tiny$\mathcal{B}$}})}{j'}+\frac{C_{\mbox{\tiny$\mathcal{B}$}}m}{a}-\frac{m^2}{\pi}(\ln2+1/2)\right].
\end{eqnarray}
To compute the radiative correction to the Casimir energy for the case of MBC, we begin by utilizing Eqs. (\ref{Greens.function.MBC.without.LV&Roughness}) and (\ref{Vacuum.En.EXP2.}). By performing the integration over momentum and spatial coordinates, and applying the cutoff regularization technique as done in Eq. (\ref{cutoff.Reg.}), we obtain:
\begin{eqnarray}\label{Equation.MBC.VAC.}
      \Delta E_{\mbox{\tiny Vac.$\mathcal{M}$}}^{(1)}=\frac{-\lambda L^2}{128\pi^2a}
     \sum_{n,n'=0}^{\infty}\Big(1+\frac{1}{2}\delta_{n,n'}\Big)\tilde{\Omega}_{n,\mbox{\tiny$\mathcal{M}$}}\tilde{\Omega}_{n',\mbox{\tiny$\mathcal{M}$}}+2\times\{a\to\mbox{\small$\frac{L-a}{2}$}\}-\{a\to b\}-2\times\{a\to\mbox{\small$\frac{L-b}{2}$}\},
\end{eqnarray}
where $\tilde{\Omega}_{n,\mbox{\tiny$\mathcal{M}$}}^2=(\mbox{\small$n+\frac{1}{2}$})^2\pi^2/a^2+m^2$. Next, by applying the APSF provided in Eq. (\ref{APSF.MBC}), we convert the summation forms in Eq. (\ref{Equation.MBC.VAC.}) into integral forms. So, we have
\begin{eqnarray}\label{after.APSF.MBC.RC}
         \Delta E_{\mbox{\tiny Vac.$\mathcal{M}$}}^{(1)}&=&\frac{-\lambda L^2}{128\pi^2}\left[\frac{m^4a}{\pi^2}\left(\int_{0}^{\infty}d\xi\sqrt{\xi^2+1}\right)^2
        +\frac{m^3}{2\pi}\int_{0}^{\infty}d\xi\big[\xi^2+1\big]+\frac{2m^2B_1(a,m)}{\pi}\int_{0}^{\infty}d\xi\sqrt{\xi^2+1}\right.\nonumber\\&&\left.
      +\frac{B_1(a,m)^2}{a}+\frac{1}{2a}B_2(a,m)\right]
    +2\times\{a\to\mbox{\small$\frac{L-a}{2}$}\}-\{a\to b\}-2\times\{a\to\mbox{\small$\frac{L-b}{2}$}\}.
\end{eqnarray}
The divergences arising from the first three terms of Eq. (\ref{after.APSF.MBC.RC}) are similar to those observed in Eq. (\ref{Subtraction.Vac.4}). For instance, by applying the BSS to the first term on the right-hand side of Eq. (\ref{after.APSF.MBC.RC}), similar to what was done in Eq. (\ref{remove.BSS.1}), all its infinities are eliminated. The second term on the right-hand side of Eq. (\ref{after.APSF.MBC.RC}) is independent of \(a\) and \(b\), so its contribution is automatically cancelled during the subtraction process provided by the BSS. The finite part remaining from the third term on the right-hand side of Eq. (\ref{after.APSF.MBC.RC}) was derived in Eq. (\ref{remove.BSS.22}). As a result, for Eq. (\ref{after.APSF.MBC.RC}), we obtain:
\begin{eqnarray}\label{after.cancel.MBC.RC}
       \Delta E_{\mbox{\tiny Vac.$\mathcal{M}$}}^{(1)}=\frac{-\lambda L^2}{128\pi^2a}\left[B_1(a,m)+\big(\mbox{$\frac{1}{2}$}+\ln2\big)\right]B_1(a,m)
     +2\times\{a\to\mbox{\small$\frac{L-a}{2}$}\}-\{a\to b\}-2\times\{a\to\mbox{\small$\frac{L-b}{2}$}\},
\end{eqnarray}
where $B_2(a,m)=0$ and the branchcut $B_1(a,m)$ is:
\begin{eqnarray}\label{Branchcut.MBC.RC.}
           B_1(a,m)=\frac{2m^2a}{\pi}\int_{1}^{\infty}\frac{(\eta^2-1)^{\frac{1}{2}}}{e^{2ma\eta}+1}d\eta
          =\frac{-m}{\pi}\sum_{j=1}^{\infty}\frac{(-1)^jK_1(2maj)}{j}.
\end{eqnarray}
The final step of the computation involves evaluating the limits specified in Eq. (\ref{BSS.DEF}). After completing this step, the final result for the radiative correction to the Casimir energy of a massive self-interacting scalar field confined by MBC between two parallel plates is obtained as:
\begin{eqnarray}\label{Final.ECas.RC.MBC}
      \mathcal{E}_{\mbox{\tiny Cas.$\mathcal{M}$}}^{\mbox{\tiny (1)}}(m,a)=\frac{-\lambda L^2m}{128\pi^3}\sum_{j=1}^{\infty}\frac{(-1)^jK_1(2maj)}{j}\left[\frac{m}{\pi a}\sum_{j'=1}^{\infty}\frac{(-1)^jK_1(2maj')}{j'}-\frac{m^2}{\pi}(\ln2+1/2)\right]
\end{eqnarray}

\begin{acknowledgments}
The author would like to thank the Research Office of Semnan Branch, Islamic Azad University, for their financial support.
\end{acknowledgments}
%\bibliography{apssamp}

\end{document}